\documentclass{appolb}
\usepackage{epsfig}

\begin{document}
\title{Photoproduction of $\eta$-mesons from light nuclei
}
\author{B. Krusche\\
for the TAPS- and A2-collaborations
\address{Department of Physics and Astronomy, University of Basel,\\
CH-4056 Basel, Switzerland}
}
\maketitle
\begin{abstract}
\section*{Abstract} 
In a series of experiments coherent and quasifree $\eta$-photoproduction
from light nuclei ($^4$He, $^3$He, $^2$H) was investigated
with the TAPS-detector at the Mainz MAMI-accelerator. 
The experiments were motivated by two different subjects: the determination of 
the isospin structure of the electromagnetic excitation of the S$_{11}$(1535) 
resonance and the study of the $\eta$-nucleon and $\eta$-nucleus interaction 
at small momenta. The results for the deuteron and $^4$He are summarized and 
first preliminary results for $^3$He are presented. 
\end{abstract}
  
\section{Introduction}

Photoproduction of mesons from light nuclei can serve two
purposes: the investigation of the isospin structure of the electromagnetic
excitation of nucleon resonances and the study of the meson - nucleon 
interaction via final state interaction effects. In principle, there are two 
possibilities to learn about the isospin structure of the photoexcitation 
amplitudes. Coherent photoproduction from light nuclei may be used as an 
isospin filter, while photoproduction from bound nucleons in quasifree 
kinematics can be used to extract the neutron cross section. The small binding 
energy and the comparatively well understood nuclear structure single out 
the deuteron as exceptionally important target nucleus. However, as a
cross check for nuclear effects, it is desirable to use also a nucleus
with a different momentum distribution of the bound nucleons as target. 
In this respect $^4He$ is an extreme case due to the stoung bindung of 
the nucleus.

During the last few years we have intensively studied quasifree, and coherent
$\eta$-photoproduction from the deuteron and from He-nuclei 
\cite{Krusche_1}-\cite{Weiss_2}. 
The electromagnetic excitation of $I=1/2$ $N^{\star}$ resonances involves two 
isospin components, the isoscalar and the isovector amplitudes. The complete 
determination of the amplitudes is possible from the measurement of 
$\eta$-photoproduction from the proton, the neutron and coherent 
photoproduction from the $I=0, J=1$ deuteron making use of:

\begin{equation}
  \sigma_p \sim |A^{IS}_{1/2} + A^{IV}_{1/2}|^2,\;\;\;\;\;
  \sigma_n  \sim  |A^{IS}_{1/2} - A^{IV}_{1/2}|^2,\;\;\;\;\;
  \sigma_d  \sim  |A^{IS}_{1/2}|^2
\end{equation}

where $A^{IS}_{1/2}$ denotes the isoscalar and $A^{IV}_{1/2}$ the isovector 
part of the helicity amplitude.  It is of course clear that the extraction of 
the amplitudes from the cross sections requires a careful separation of 
the contributions from overlapping resonances and background terms like 
Born terms or vector meson exchange. The situation is simplest close to 
threshold where only a few partial waves contribute. In this sense 
$\eta$-photoproduction in the threshold region is a special case, since the
reaction is completely dominated by the excitation of the S$_{11}$(1535)
resonance.

\section{Experiments}
The experiments were done at the Mainz MAMI accelerator with the TAPS-detector
\cite{Novotny,Gabler}, experimental details can be found in 
\cite{Krusche_1}-\cite{Hejny_2}.
The extraction of the cross section ratio $\sigma_n/\sigma_p$ for the reaction
on the neutron and the proton can be done with different experimental 
concepts, which have all been exploited.
\begin{itemize}
\item{Measurement of the inclusive $d(\gamma ,\eta)X$ reaction and comparison 
of the sum of Fermi smeared proton cross section and Fermi smeared ansatz for 
the neutron cross section to the inclusive nuclear cross section in PWIA.
Variation of $\sigma_n$ until agreement is achieved.}
\item{Coincident measurement of $\eta$-mesons and recoil nucleons. Extraction
of the ratio of $\sigma_n /\sigma_p$ as function of the incident photon
energy $E_{\gamma}$.}
\item{Coincident measurement of $\eta$-mesons and recoil nucleons and 
reconstruction of the effective $\sqrt{s^{\star}}$ and effective
incident photon energy $E_{\gamma}^{\star}$ from the final state kinematics,
extraction of $\sigma_n /\sigma_p$ as function of 
$E_{\gamma}^{\star}$:
\begin{equation}
s^{\star}=(E_{\eta}+E_p)^2-(\vec{p}_{\eta}+\vec{p}_R)^2
\end{equation}
\begin{equation}
E_{\gamma}^{\star}=\frac{s^{\star}-m_R^2}{2m_R}\;\;.
\end{equation}
}
\end{itemize}

\section{Results}
\subsection{Inclusive quasifree $\eta$-production}
\begin{figure}[hbt]
\label{fig:incl}
\begin{minipage}{0.cm}
{\mbox{\epsfysize=6.cm \epsffile{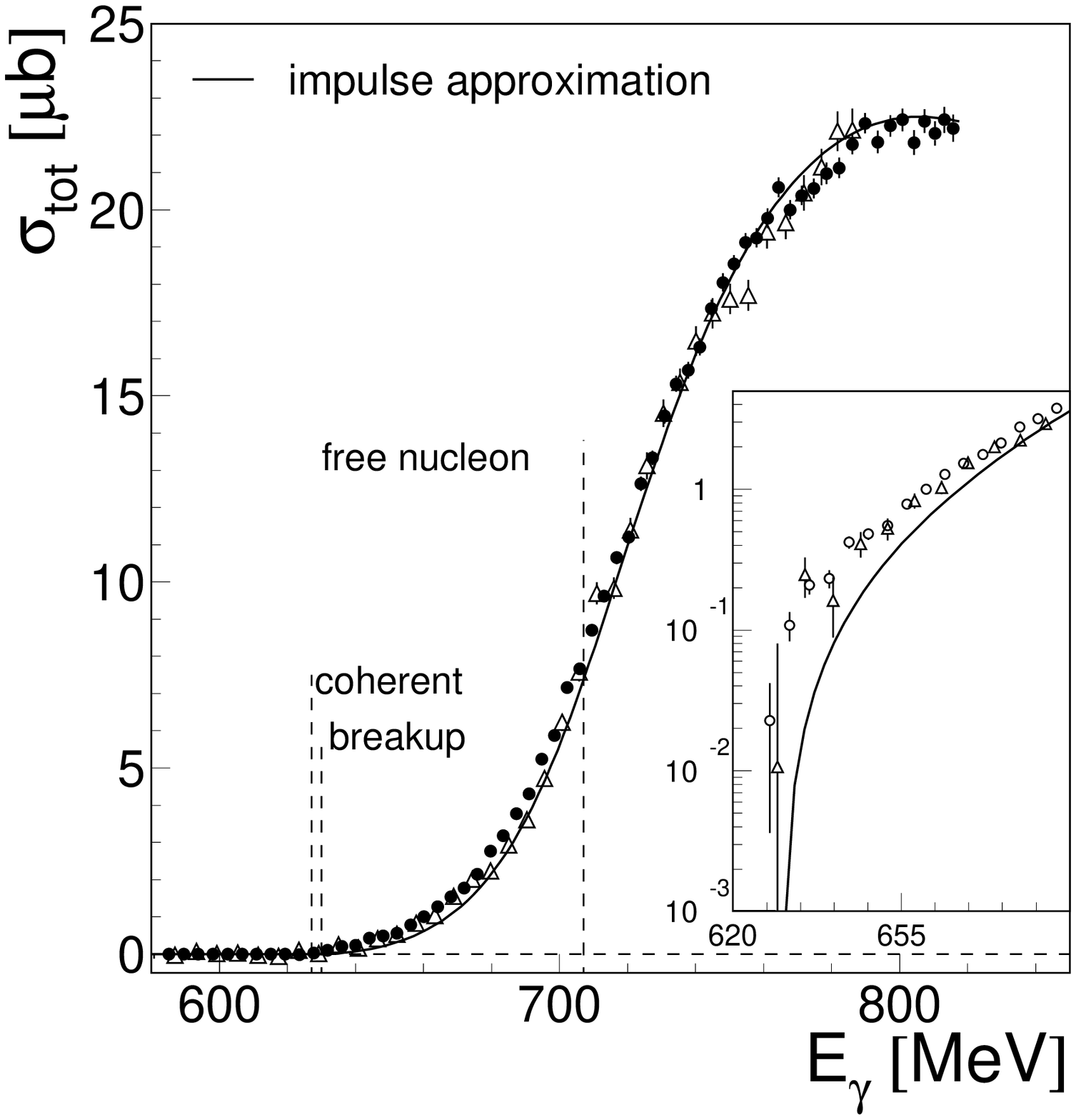}}}
\end{minipage}
\hspace{6cm}
\begin{minipage}{0.cm}
{\mbox{\epsfysize=6.cm \epsffile{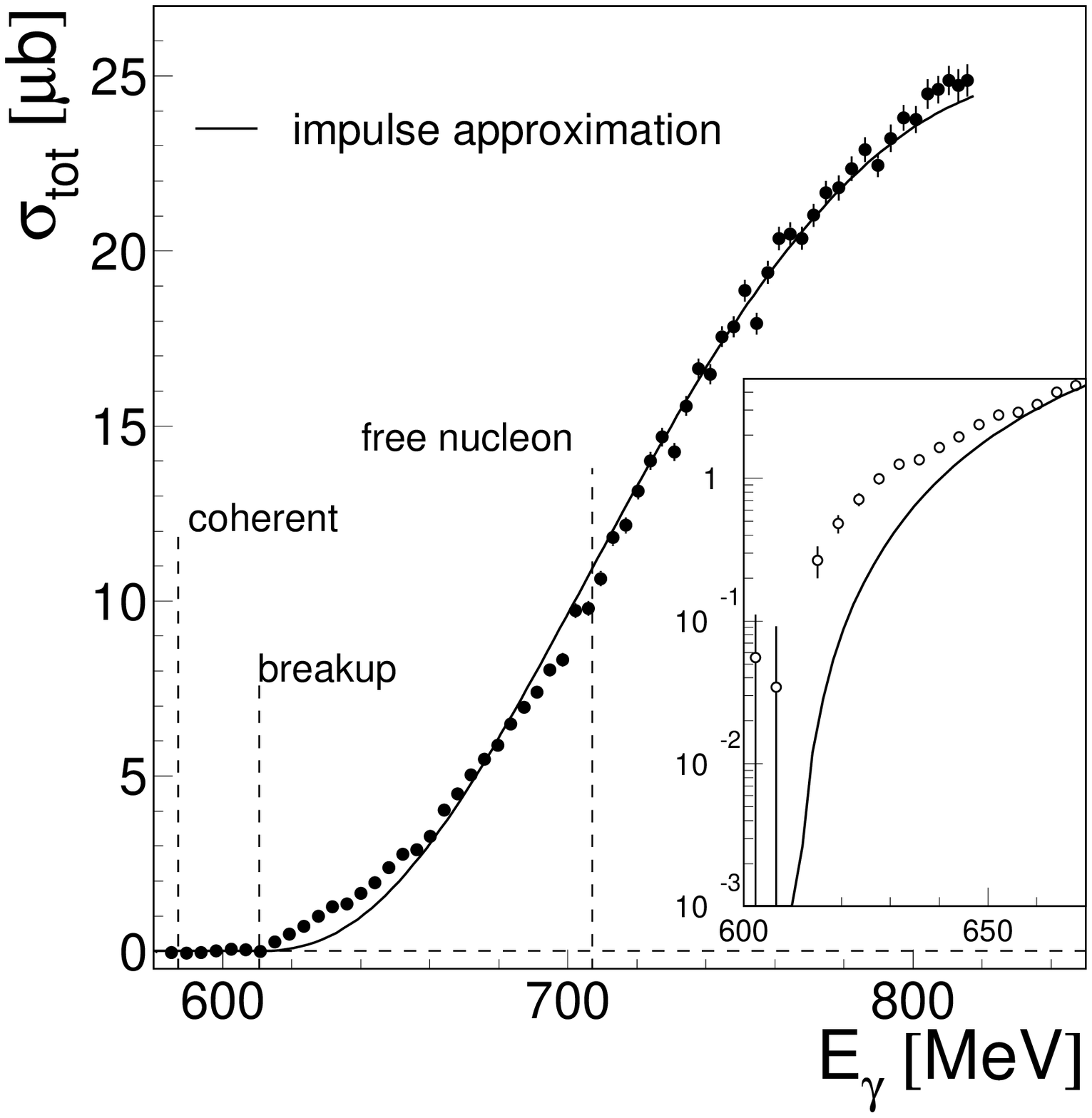}}}
\end{minipage}
\caption{Inclusive $\eta$ photoproduction cross section. Left side:
from the deuteron. Circles: ref. \cite{Weiss_2}, triangles: ref. 
\cite{Krusche_1}. Right side: from $^4$He \cite{Hejny_1}. For both pictures 
the dashed lines indicate the coherent, the breakup and the 
free nucleon production thresholds. The solid curves are the result of the 
impulse approximation model under the assumption of a constant
$\sigma_n /\sigma_p$=2/3 ratio (see text). Inserts: threshold region. 
}
\label{fig:tot_incl}       
\end{figure}

The total cross sections for the reactions $d(\gamma ,\eta)X$ and 
$^4He(\gamma ,\eta)X$ are summarized in fig. \ref{fig:incl}.
The result of the PWIA analysis in the S$_{11}$(1535) range
was a constant ratio $\sigma_n /\sigma_p$=2/3 \cite{Krusche_1,Hejny_2}.
Only very close to the threshold the agreement between PWIA and experimental 
results is not good (see inserts in fig. \ref{fig:tot_incl}). This is due to 
final state interaction (FSI) effects which have been understood in the 
meantime \cite{Elster}.  
Note that the influence of the different nuclear momentum distributions is 
quite severe. The total cross sections from $^4He$ and from the deuteron 
are almost equal at an incident photon energy of 800 MeV although twice 
the number of nucleons is involved in the He case. The good agreement of 
both data sets with impulse approximations using the same neutron
proton ratio is therefore quite reassuring for the application of this method.

\subsection{Exclusive quasifree $\eta$-production}
\begin{figure}[hb]
\begin{center}
\epsfysize=7.cm \epsffile{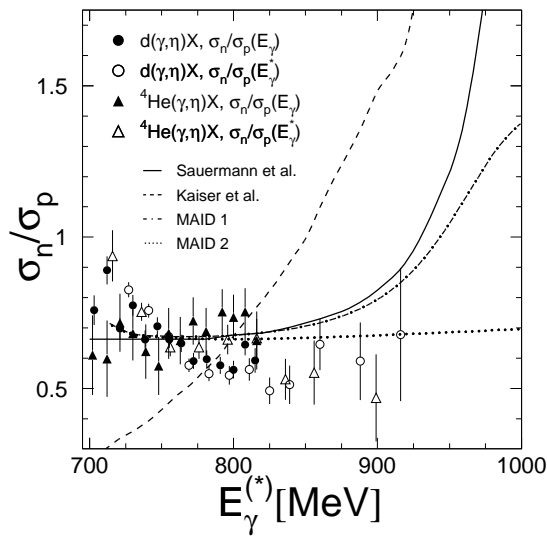}
\caption{Ratio of exclusive proton and neutron cross sections in the TAPS
acceptance. The deuteron data are compared to the $^4He$-data \cite{Hejny_1}
and to model predictions  from Sauermann et al. \cite{Sauermann} and
Kaiser et al. \cite{Kaiser}. The curves labeled MAID are the 
predictions from the MAID model \cite{MAID} for the full model (MAID1) and
restricted to the contributions from the S$_{11}$(1535)-resonance, Born terms
and vector meson exchange (MAID2).}
\label{fig:ratall}       
\end{center}
\end{figure}

The data for the neutron - proton cross section ratio obtained with the
exclusive measurements for different targets are summarized in fig. 
\ref{fig:ratall} and compared to model predictions.
The observed rise of the cross section ratio to unity at the breakup threshold 
for incident photon energies, respectively at the free nucleon threshold for
equivalent photon energies, is understood and a pure nuclear effect due to 
rescattering contributions etc. . The ratios are almost constant for higher 
incident photon energies and within their systematical uncertainties 
consistent with $\sigma_n /\sigma_p$=2/3. 
A comparison of the energy dependence in the S$_{11}$-range to model 
predictions clearly disfavors the interpretation of the resonance as a 
$K\Sigma$ bound state \cite{Kaiser}. 
\vspace*{-1.cm}

\subsection{Coherent $\eta$-production}
\begin{figure}[hb]
\begin{center}
\epsfysize=5.5cm \epsffile{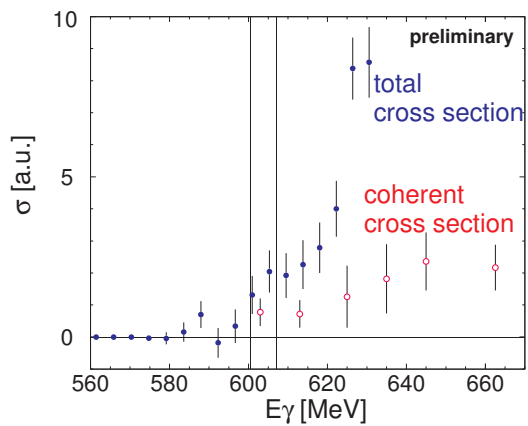}
\caption{Preliminary results for the $^3He(\gamma ,\eta)X$ reaction 
(full symbols) and the $^3He(\gamma ,\eta)^3He$ reaction (open symbols) 
\cite{Pfeiffer}.}
\label{fig:marco}       
\end{center}
\end{figure}

The results from the quasifree $\eta$-photoproduction in combination with
coherent production from the deuteron \cite{Weiss_1} have been interpreted as 
evidence for a dominant isovector contribution in the photoexcitation of the
S$_{11}$-resonance. Since in addition the relevant multipole ($E_{o^+}$) is a
spin-flip multipole it is expected that the coherent reaction on $^4He$
($I=0$, $J=0$) is almost completely supressed, that it is weak on the deuteron
($I=0$, $J=1$) and not suppressed on $^3He$ ($I=1/2$, $J=1/2$). 
In agreement with this expectation, only upper limits could be derived for the
coherent reaction on $^4He$ \cite{Hejny_1}, while a small contribution
was seen for the deuteron \cite{Weiss_1}. However, in case of the deuteron
the results for the quasifree and coherent reaction are not yet completely
understood since the first seem to indicate  an isoscalar admisture of 
$\approx$ 9\% while the latter require a larger isoscalar contribution
of $\approx$ 20\% \cite{Weiss_1}. The problem could be due to nuclear effects
in the coherent reaction which are not yet well enough under control.
Additional information will be gained from the coherent reaction on $^3He$,
where due to the quantum numbers of the nucleus the dominant isovector,
spin-flip part of the amplitude contributes. First preliminary results for the
the cross section of this reaction indicate indeed a substantial contribution
from the coherent part (see fig. \ref{fig:marco}).  

\vspace*{-1cm}

\end{document}